\documentclass[12pt,letter]{emulateapj}
\usepackage{amsmath,multirow,dcolumn,fancyhdr,charter,graphicx, tabularx}
\usepackage{graphics}
\usepackage{color,ulem,epstopdf}

\newcommand{\rearth}{$R_{\oplus}$}

\newcommand{\kep}{{\it Kepler}}

\newcommand{\ee}{$\eta_{\oplus}$}
\def\refnew#1{(\ref{#1})}

\shorttitle{A Statistical Reconstruction of the Planet Population around \kep{} Solar-Type Stars}
\shortauthors{Silburt et al.}

\begin{document}

 
\thispagestyle{plain}


\title{A Statistical Reconstruction of the Planet Population around \kep{} Solar-Type Stars}
\author{Ari Silburt$^{1}$, Eric Gaidos$^{2}$, \& Yanqin Wu$^{1}$}
\affil{$^1$Department of Astronomy and Astrophysics, University of
  Toronto, Toronto, ON M5S 3H4, Canada;} \affil{$^2$Department of
  Geology \& Geophysics, University of Hawai`i at M\={a}noa, Honolulu,
  HI 96822 USA}


\begin{abstract} 
Using the cumulative catalog of planets detected by the
NASA {\it Kepler} mission, we reconstruct the intrinsic occurrence of Earth- to
Neptune-size (1 -- 4\rearth{}) planets and their distributions with radius and orbital
period.  We analyze 76,711 solar-type ($0.8<R_*/R_{\odot}<1.2 $) stars with 430 planets 
on 20--200~d orbits, excluding close-in planets that may have been affected by
the proximity to the host star.  Our analysis considers errors in
planet radii and includes an ``iterative simulation'' technique that
does not bin the data.  We find a radius distribution that peaks at
2--2.8 Earth radii, with lower numbers of smaller and larger planets.
These planets are uniformly distributed with logarithmic period, and
the mean number of such planets per star is $0.46 \pm 0.03$.  The
occurrence is $\sim 0.66$ if planets interior to 20~d are included.  We
estimate the occurrence of Earth-size planets in the ``habitable zone" (defined as
1--2\rearth{}, 0.99--1.7 AU for solar-like stars)
as $6.4^{+3.4}_{-1.1} \%$.  Our results largely agree with those of \cite{Petigura2013}, 
although we find a higher occurrence of 2.8--4
Earth-radii planets. The reasons for this excess are the inclusion of errors in 
planet radius, updated \cite{Huber2014} stellar parameters, and also the 
exclusion of planets which may have been affected by proximity to the host star.
\end{abstract}

\section{Introduction}
\label{sec:introduction}

Anaximander of Miletus proposed that there were many Earth-like worlds
\footnote{{\it Dissertation on the Philosophy of Aristotle, in which
    his Principal Physical and Metaphysical Dogmas are Unfolded},
  transl. by T. Taylor, London, 1812}.  In the twenty five centuries
since the Ionian philosopher's speculation, we have seen an
accelerating convergence towards an answer: many if not most stars
have planets, Earth-sized (and presumably rocky) bodies are more 
common than Jupiter-sized gaseous bodies, and
some Earth-sized planets are on orbits in the so-called
``habitable zones'' of their stars \citep{Schneider2011}. The most recent advances in this
field have come from data collected by the \kep{} transiting-planet
mission \citep{Borucki2010}.  Four years of observations of a $\sim
100$ sq. deg. field have led to the identification thus far
of 4234 confirmed or candidate transiting planets, about 84\% of the
candidates appear to be smaller than Neptune (3.88\rearth{}) \citep{NASAEA}.

Beyond simple human curiosity about the prevalence of Earth-like
planets and possibility of life elsewhere, planet statistics provide
an important test of planet formation models \citep[e.g.][]{Benz2014}.
For example, the distribution of planets with respect to orbital
period and mean motion resonances test models of planet formation and
early migration \citep[e.g.,][]{HansenMurray,Baruteau2013}.  Planet
radius and period distributions can be combined to reconstruct the
distribution of solid mass in disks
\citep[e.g.,][]{Chiang2013,Raymond2014}.

There have been many previous studies that use the \kep{} data to
infer the intrinsic population of planets around \kep{} target stars,
or subsets of those stars
\citep[e.g.][]{Catanzarite,Youdin,Traub2012,Howard2012,Fressin2013,
  Dressing2013,Gaidos2013,Dong2013,Kopparapu2013b,Petigura2013}.
These works differ in their samples and methods, but are broadly
consistent in estimating that the occurrence of planets per star on orbits of days to months
to be of order unity.  Some of these works have estimated \ee{}, the
occurence of Earth-size planets in a circumstellar ``habitable zone''
(where stellar irradiation is similar to the solar constant), and find
that \ee{} is of order tens of percent.  Of particular interest to us
is the work of \citet[][hereafter PHM13]{Petigura2013} who performed
an independent analysis of the \kep{} photometric lightcurves, both
identifying candidate planet transits and determining the detection
efficiency by injecting synthetic transit signals into \kep{}
lightcurves and recovering them.  Among the salient conclusions from
PHM13 are that the distribution of planets peaks at a radius of
2--2.8\rearth{} and that, \ee{} is about 22\% (using a liberal
definition of the habitable zone).

We identify two reasons to revisit the derivation of the \kep{} planet
population. Firstly, we want to more fully account for the
uncertainties and biases in the \kep{} data and related observations
of the target stars.  Secondly, we wish to consider a ``primordial''
planet population, and restrict our analysis to planets far enough
from their host stars such that their properties have not been altered by
proximity since their formation.

Precise statements on the occurrence of planets requires rigorous
statistical methods, full accounting of errors, and adequate
assessment of potential biases.  First, while the overall rate of
``false positives'' among \kep{} candidate planets appears to be low
\citep{Morton2011,Fressin2013}, it is not uniform across all
  periods and all sizes \citep{Santerne2013}.  Second, determination
of planet occurrence from transit surveys requires accurate estimates
of detection efficiency (also known as ``completeness''), which depends on the
parameters (i.e. density and/or radius) of not only the planet host
stars but of the entire target catalog. The parameters of \kep{} stars
were first determined by combining multi-wavelength photometry,
stellar models, and Bayesian inferrence \citep{Brown2011}.  Colors of
solar-type stars depend only weakly on gravity and metallicity, and
parameter values based on photometry have large random and systematic
errors in both effective temperature \citep{Pinsonneault2012}, gravity
and luminosity class \citep{Mann2012}.  Spectroscopy, asteroseismology,
and improved stellar models have yielded more reliable parameters 
\citep{Huber2014}, especially for \kep{} planet-hosting stars 
(\kep{} Objects of Interest or KOIs).  Nevertheless, 70\% of all stars in the
\citet{Huber2014} catalog have assigned parameters based on KIC
photometry; the median upper and lower fractional errors in stellar
radius among these solar-type stars is 40\% and 10\%, respectively. Because
the estimated radius of a planet detected by transit depends on the
stellar radius, and the probability of transit depends on stellar
density, these errors need to be taken into account when computing
occurrence rates.  Errors in luminosity also affect the certainty with
which a planet can be assigned to a habitable zone described by a
range of stellar irradiance \citep{Gaidos2013,Mann2013}.

Third, biases, if uncorrected or unaccounted for, will distort our 
perspective on planet populations.  
\citet{Mann2012} found that up to $96\%$ of
the reddest \kep{} target stars are giants, even though virtually all
red KOI hosts are dwarfs for the simple reason that it is extremely
difficult to detect a transiting planet around a giant star.  They
showed that dilution of the target catalog by giants had led to an
underestimate of the occurrence rate and an incorrect claim that M
dwarf hosts of detected planets are redder and thus more metal-rich
than those without detected planets.

\citet{GaidosMann2013} showed that because the \kep{} target catalog
is essentially magnitude-limited, Malmquist bias combined with
uncertainties in stellar parameters means that stellar distances are
underestimated and many stars are likely to be more luminous, evolved,
and larger than their nominal values.  Follow-up observations and
analysis thus far seem to confirm this
\citep[e.g.][]{Bastien2014,Everett2013,Verner2011}.  Because
transiting planet radius scales with increasing stellar radius, this means that
planet radii are underestimated.  Moreover, the rate of planet
detection decreases with increasing stellar radius or density, thus
for a given planet radius, the detection rate is overestimated and
thus the occurrence is underestimated.  Detection bias again means
that any estimate of this effect based on the host stars of transiting
planets is an {\it underestimate}: the effect will be greater among
stars in the overall target catalog.  Another bias is Eddington bias:
scatter by error from more populated regions of parameter space into
less populated regions will produces the opposite effect,
i.e. occurrence will be overestimated \citep{GaidosMann2013}.

Previous analyses of the \kep{} planet population have sometimes not taken
these errors or biases into account, and instead have considered only
Poisson (counting) statistics \citep[e.g.][]{Petigura2013,Howard2012}.
Finally, many analyses
were performed by binning the data into discrete bins of planet radius $R_p$
and orbital period $P$.  While simple and readily explicable, the binning
method runs the risk of masking details of a distribution, especially
that of radius, which may be important for testing theoretical
models.

In this work consider a ``primordial'' population of
planets, as opposed to one that has evolved under the influence of the
host star. Effects of the latter, including tidal heating
\citep{Jackson2008}, atmospheric escape \citep{Tian2005}, ohmic heating
\citep{Batygin2011}, and impact erosion \citep{Marcus2009}, act with
an efficiency that is inversely proportional to the distance to the
host star. In particular, \citet{Owen2013} proposed that
photoevaporation by stellar X-ray and ultraviolet irradiation have effectively removed
the hydrogen envelopes of close-in planets ($P \leq 10$~d), leading
to the observed paucity of super-Earth sized planets in that
neighbourhood.  This process was also investigated for a few \kep{}
systems by \citet{Lopez2012}. Regardless of the mechanism, the
distinctiveness of the $P< 20$~d and $P>20$~d populations \citep[see,
e.g.][]{Youdin} suggests that any analysis treat these separately.  In
this study, we focus exclusively on the latter population as we
believe it is more likely to represent the ``primordial'' state.  On
the other hand, because of \kep{}'s low efficiency at detecting
long-period planets (see \S \ref{sec:completeness}), we are forced to
limit our consideration to planets with $P<200$d.

Practical reasons also limit the range of planet radius $R_p$
considered.  Although \kep{} can readily detect a transiting giant
planet, the occurrence of these objects is indubitably much lower than
that of smaller planets.  The distribution with planet radius falls to
a very low level beyond Neptune-size objects: only 8\% of \kep{}
candidate planets have nominal radii $>8$\rearth{}, and the
false-positive rate increases as well \citep{Santerne2012,Colon2012}.
Conversely, \kep{} can detect planets smaller than 1\rearth{} for only
a tiny fraction of stars, mostly M dwarfs.  For these reasons we
restrict our analysis to a radius range of 1--4\rearth{} over which
statistically rigorous analyses can be performed. 

In this contribution, we infer the intrinsic distribution of planets
with $20<P<200$~d (equivalent to 0.16--0.67~AU) and
$1 R_\oplus < R_p < 4R_\oplus$ around solar-type stars as
observed by \kep{} over
its entire mission (Quarters 1--16).  This analysis includes the
effects of errors in stellar and planet radius, and takes into account
some of the biases that may affect previous works.  We introduce a
method of iterative simulation to determine the radius distribution of
planets without resorting to binning.  We compare our results with
those of PHM13 and also carry out a detailed comparison of the two
methods to understand the source of any discrepancies.  Finally, we use our
simulations to assess the effect of systematic bias, namely an overall 
underestimate of stellar radius, on
inferences of a planet population from the \kep{} catalog.

\section{Methods}
\label{sec:methods}
\subsection{Catalogs of Stars and Planets}
\label{sec:catalogs}
We construct a stellar sample from the \cite{Huber2014} catalog of
196,468 stars observed during the \kep{} mission (Quarters 1--16),
selecting ``Sun-like'' stars with radii $0.8R_{\odot}<R_*<1.2R_{\odot}$. We
restrict the sample to stars with a \kep{} magnitude $K_{p}<15.5$ to
avoid faint stars with uncertain properties and noisy lightcurves.
This leaves a sample of 76,711 solar-type stars, hereafter known as
the ``Solar76k'' sample.  Our planet sample is constructed from the 26
February 2014 version of the KOI catalog \citep{Ramirez2014}.  Planet
and stellar parameters are updated with values from \cite{Huber2014}
where available.  In addition to the cuts made on our stellar sample
we also require $20<P<200$ d, $0.5R_{\oplus}<R_p<6R_{\oplus}$ and
SNR$>$12. Although we are only interested in the occurrence of planets 
$1R_{\oplus}<R_p<4R_{\oplus}$, we use a larger radius range for our analysis 
since these planets have a non-zero probability of being in our region of 
interest after accounting for radius errors (\S\ref{sec:plerr}).
This leaves us with 430 candidate planets, hereafter known
as the ``430KOI'' sample. The full 430KOI dataset is published in the
electronic edition of the {\it Astrophysical Journal}. We include a partial table (Table~\ref{tab:430KOI}) 
at the end of this paper for guidance regarding its form and content.

We also construct a second planet sample
(used only in \S\ref{sec:intperiod}) using the same cuts above
except relaxing the period restriction to $5<P<200$.  This leaves us with
1052 KOIs, hereafter known as the ``1052KOI'' sample.

To ease comparison with PHM13, we retrieve their vetted sample of
603 planet candidates that fall within $5<P<100$~d. We update the stellar
parameters where possible, using \citet{Huber2014}.  We hereafter
refer to this planet sample as the ``603PHM'' sample.

\subsection{Simulated Planet Detections}
\label{sec:geneq}

Our simulator synthesizes single planet-star pairs\footnote{ We
  ignore the occurrence of multiple systems and assume that each
  planet has an independent occurrence.},  drawing stars from one of
the catalogs described above, and planet parameters from a large
``master'' population as we describe in \S \ref{sec:plerr}.  We calculate whether each
planet transits its host star in a probabilistic manner, and then determine
whether \kep{} could have detected it.  We compare the
properties of these simulated detections with the observed candidate \kep{}
planets, and modify the master population using two different
techniques: iterative Monte Carlo Markov Chain (MCMC) and Iterative Simulation
(IS).  In implementing our simulations we make two critical
assumptions. First, that the orbital periods and radii of \kep{}
planets (as well as orbital eccentricities), are independently
distributed, i.e. the occurrence is a separable function of period and
radius:
\begin{equation}
{{d^2 N}\over{{d\log P \, d\log R}}} = p(P) r(R)\, ,
\label{eq:assumption}
\end{equation}
where $p$ and $r$ are some yet-to-be-determined functions, and  
$N$ is the total number of planets. 
Second, we assume that these distributions do not vary over the range
of stellar parameters considered.  We discuss these assumptions in \S
\ref{sec:sensitivity}.  In this study, we further specify that the
  period distribution is a power-law:
\begin{equation}
\frac{dN}{d log P} = C P^{\alpha}\, ,
\label{eq:P}
\end{equation}
where $C$ is a normalization constant.

The geometric probability that a planet transits its host star is
\citep{Winn},
\begin{equation}
\label{eqn.probtransit}
p = \frac{R_*}{a}\frac{1+e \sin \omega}{1-e^2},
\end{equation}
where $a$ is the semimajor axis, $e$ the orbital eccentricity, and
$\omega$ the argument of periastron.  While $a$ can be calculated from
$P$ and the estimated mass of the host star, the orbital eccentricity
of \kep{} planets are unknown and must be estimated statistically.
Assuming a Rayleigh distribution with
  dispersion $\sigma_e$, \citet{Moorhead2011} estimated $\sigma_e =
0.2$ by studying the distribution of transit durations. This
  is likely affected by uncertain stellar radii and may be an
  overestimate. TTV studies have led to much smaller eccentricity
  dispersion ($\sigma_e \sim$ a few percent), at least in multiple
  planet systems \citep{WuLithwick,HaddenLithwick}. Here, we choose
$\sigma_e = 0.18$ and show that our results are not sensitive to the
exact value of $\sigma_e$ (\S \ref{sec:sensitivity}). 
The underlying distribution of $\omega$ can be safely assumed to be
uniform over [0,$2\pi$].  
Integrating $p$ over the distributions of $e$ and $\omega$,
Eq.\refnew{eqn.probtransit} becomes $p = p_0 R_*/a$, with
$p_0=1.073$.  As in previous works, we require that at least three
transits have been observed. 
We scale every transit probability by $(1/p_0)(a/R_*)_{max}$, 
the inverse of the max transit probability. Since transiting planets
occur only for $<5$ degrees of inclination, this scaling is used to speed up
the rate of transiting planets. Otherwise, we would have to wait long 
periods of time in order acquire the large numbers of transiting planets we require
to conduct this analysis. 

We now proceed to assign a transit duration, $T$, to a given
transiting planet. We follow the procedure of \citet{Gaidos2013} 
by setting
\begin{equation}
\label{eq:T_IS}
T = \tau^{2/3}P^{1/3}\Delta\, ,
\end{equation}
where $\tau = 2\sqrt{R_*^3/(\pi G M_*)}$ is the stellar free-fall
time, $G$ the gravitational constant, $M_*$ the stellar mass, and
\begin{equation}
\label{eqn.delta}
\Delta = \frac{\sqrt{(1-e^2)(1-b^2)}}{1 + e \cos \omega}\, , 
\end{equation}
with $b$ being the impact parameter. For $a \gg R_*$, the impact
parameter $b$ is uniformly distributed in the range [0,1].
We then calculate $dN/d\Delta$, the likelihood of drawing a
given $\Delta$, or rather, its cumulative distribution,
$\overline{N(\Delta)} \equiv \overline{\int_0^\Delta
{dN}\over{d\Delta'}}d\Delta'$, with the overbar indicating 
marginalization over $e$ and $\omega$. 
Using the chain rule, we find
\begin{equation}
\label{eq:delta1}
{\frac{dN}{d\Delta}} d\Delta = 
{\frac{dN}{db}}
{\frac{db}{d\Delta}} d\Delta = 
{\frac{db}{d\Delta}} d\Delta,
\end{equation}
where we have used the fact that $dN/db = 1$ (i.e. $b$ is uniformly distributed) 
for transiting systems.
As a result, $\overline{N(\Delta)} = \overline{b} = \overline{\int_0^\Delta
  {{db}\over{d\Delta'}} d\Delta'}$.  Inverting
Eqn. \refnew{eqn.delta} then yields:
\begin{equation}
\overline{N(\Delta)} = \int_0^1 \eta(e) de \int_0^{2\pi} \sqrt{1 -
  \frac{\Delta^2\left(1 + e \cos \omega\right)^2}{\left(1-e^2\right)}}d\omega, 
\end{equation}
where $\eta(e)$ is the assumed eccentricity distribution.  

We also need to assign a radius to each trial planet: this
process differs between our MCMC and IS methods and is described in
their respective sections. Moreover, in comparing the radius
distribution of trial planets to the observations, we must take into
account significant uncertanties in the radius of KOIs.
We describe how we do this in the next section.

Our detection criterion is based on a comparison
between the transit signal, $(R_p/R_*)^2$, and the
effective noise over the transit duration.  \citet{Fressin2013}
established that at signal-to-noise SNR $>12$ the false-positive rate
among \kep{} KOIs is very low.  PHM13 used this criterion for their
analysis and we follow suit.  Noise in \kep{}
lightcurves is derived from photon (shot) noise, measurement error 
 (e.g. pointing error and instrument noise)
and stellar variability \citep{Koch2010}.  The \kep{} team encapsulates
the total noise of each star into quarterly transit durations of 
3-hr, 6-hr and 12-hr, known as ``CDPP''
\citep[Combined Differential Photometric Precision,][]{Christiansen2012} 
values. For a given star in a given quarter, we generate the appropriate
noise for transit duration $T$, by interpolating among the various
CDPP values using a power-law relation.  Because sources of noise
(e.g., stellar variability) are not neccesarily ``white'', 
the power-law index can and often does depart from -0.5, the
white noise value.

 We then calculate the total SNR of a model star-planet pair as
\begin{equation}
SNR = \left(\frac{R_p}{R_*}\right)^2 \left[\sum\limits_{j=1}^{16}
\frac{n_j}{\left({\rm CDPP}_j\right)^2} \right]^{1/2},
\label{eq:SNR}
\end{equation}
where $n_j$ is the number of transits in quarter $j$, and CDPP$_j$ is
the interpolated CDPP value for that quarter.  $n_j$ is found by
first assigning a randomly drawn phase for each planet, and then counting 
the number of transits in each quarter.  The system is 
proclaimed detectable if SNR $> 12$. The SNR threshold does
not account for noise that is non-Gaussian or non-stationary on a
timescale shorter than one observing quarter (90~d).  However, the
conservative requirement that SNR $> 12$ for detection
partially addresses this limitation and we consider the possible impact 
of this simplification in \S\ref{sec:petcomp} when we compare our analysis 
to PHM13. Other non-stationary effects unaccounted for in Eq.~\ref{eq:SNR}
include thermal settling events, sudden pixel sensitivity drop offs, and cosmic rays.  
Eq.~\ref{eq:SNR} also does not account for gaps in the data.

\subsection{Uncertainties in Planet Radii}
\label{sec:plerr}

As described in \S\ref{sec:introduction}, there are significant
uncertainties in the radii of most KOIs (median uncertainty = 33\%),
primarily due to our limited knowledge of the host star.  For example,
this means that there is a non-negligible chance that a planet
with a cataloged radius value of $R_p = 2.5$\rearth{}
is actually Earth-sized or Neptune-sized.  To clarify, we are not 
addressing the issue that some dwarf stars are actually giants 
(which is addressed in Section~\ref{sec:radiusdist}
when we consider that all stars are 25$\%$ larger). Instead,   
we are assuming that all claimed dwarf stars are truly dwarfs, and are 
accounting for the fact that the exact radii 
of these stars are still uncertain (on average) to $\sim 30 \%$. 

It is important that
uncertainties of such magnitude be considered, and we do this by
replacing each nominal radius by a distribution of radii governed by
Bayesian statistics.  The probability that a planet with a reported 
radius $R$ actually has a true radius
$R'$ is given by $p(R'|R) = q(R|R')r(R')$, where $r(R')$ is a
normalized prior and is the probability that a planet of radius 
$R'$ (with same period $P$) would be detected by \kep{} around a given 
star. Put another way, $r(R')$ is essentially the survey completeness 
(\S\ref{sec:completeness}) of planet $R'$ (having period $P$) with 
respect to the entire Solar76k catalog. 

In our treatment, we assume that errors in $R_*$ and hence $R_p$ are
normally distributed.  This means that
$q(R|R') = q(R'|R)$ because the Gaussian only depends on the square of
the difference $R-R'$.  We also assume that errors in stellar radius are 
uncorrelated.  This latter assumption means that a planet
with a radius that has been over/underestimated would, on average,
produce a weaker/stronger transit signal among the ensemble of target
stars and that such a planet would become less/more detectable.
If errors in $R_*$ were exactly correlated, then errors in $R_p$ would be
unaffected by considerations of detection; if all stars are smaller
then their planets will also be smaller but by the same proportion,
and thus produce transit signals of the same depth.

\begin{figure}[h]
\centerline{\includegraphics[scale=0.55]{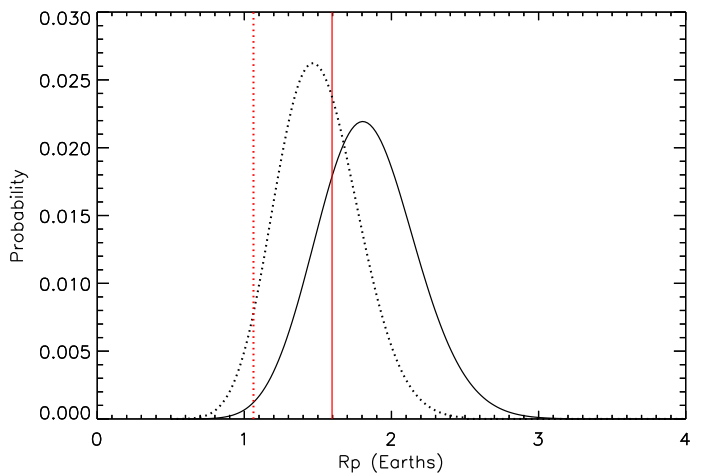}}
\caption{Example probability distributions in radius for KOI-1338.01 (solid) and 
	KOI-1925.01 (dotted). The probability distributions are shown in black, 
	while the vertical red lines represents the quoted radius 
	values. As seen in the figure, the most likely radius value can 
	differ from the catalogue value, and sometimes significantly so.}
\label{fig:prob}
\end{figure}

Provided these assumptions hold, a planet cannot be arbitrarily small,
even if the errors in radius are large, because it would never have
been detected in the first place. The $r(R')$ factor accounts for this 
fact. Our prescription for handling radius errors also accounts for 
the fact that the cataloged radius is more 
likely to be an underestimate, rather than an overestimate, of the 
true radius. This effect becomes most pronounced among KOIs with 
small cataloged radii and large uncertainty. For these cases the result is an error
distribution that is no longer a Gaussian but is strongly asymmetric,
with a cutoff just below the cataloged radius and an extended tail to
larger radii. An example probability distribution in radius
for KOI-1338.01 (solid) and KOI-1925.01 (dotted) are shown in 
Figure~\ref{fig:prob}. The vertical red lines represents the
catalogued radius values. As seen in Figure~\ref{fig:prob}, the 
most likely radius differs from the catalog value.

We implement radius errors into our analysis by replacing each KOI with a probability
distribution function (PDF) that is the product of a Gaussian times a
prior detection function which is the fraction of stars around which
the planet would be detected (i.e. completeness). We represent the 
PDF by a large number of Monte Carlo planets drawn from a Gaussian 
distribution with mean equal to the nominal value of $R_P$ and 
standard deviation equal to the cataloged error. We calculate the 
fraction of stars $F$ around which each
Monte Carlo planet could be detected.  

We then compute a normalized
CDF of $F$ with $R_p$ for each Monte Carlo set.  We can then draw a
radius value from each corrected error distribution by comparing the
CDF to a unit random deviate.  We create a ``master'' radius
distribution by randomly drawing 2 million values from all of these
distributions according to their CDFs.  We use this distribution to
represent the inferred radius distribution of observed candidate
planets after errors have been accounted for.

\subsection{Monte Carlo Markov Chain}
\label{sec:mcmc}
We implement the Monte Carlo Markov Chain according to the 
algorithm by \citet{Gelman92}.
We discretize the radius-period plane into 16 bins, 4 period bins
equally spaced in $\log P$ ($P$ = [20--40, 40--80, 80--160, 160--200]
~d) and 4 radius bins equally spaced in $\log R$ (1--1.4\rearth{},
1.4--2\rearth{}, 2-2-.8\rearth{}, 2.8--4\rearth{}). We parametrize the
planet population by a set of 4 parameters: $\alpha$ (from
Eq.\ref{eq:P}), and $\kappa_1, \kappa_2, \kappa_3$, where the
latter 3 parameters are the relative numbers of planets in the first
3 radius bins to the last bin. For each set of
parameters, we generate a mock catalog by simulating $10^5$
transiting pairs around a given stellar sample (\S \ref{sec:geneq})
while properly taking into account errors in planet radius (\S
\ref{sec:plerr}).  We then compare our mock population against
the 430KOI sample by first binning the KOIs in the same manner and then removing 
a portion from each bin to account for false positives 
\citep[Table 1 from][]{Fressin2013}. We then scale our mock catalog 
down from $10^5$ to match the total number of remaining KOIs. 
The goodness of fit is measured by comparing the simulated number of
planets in each of the 16 bins, $S_i$, versus that of the
observed, $D_i$,
\begin{equation}
\chi^2=\sum\limits_{i=1}^{16} \frac{({\rm D}_i - {\rm S}_i)^2}{{\rm S}_i}, 
\end{equation}
Here, we have assumed that the error in each bin is dominated by 
Poisson error. Our Markov chain is run for 4000 iterations, with a
``burn in'' of 200 steps that are excluded from subsequent statistical 
analysis. A new set of parameters are accepted if $\chi^2_{n} <
\chi^2_{n-1}$ (also known as Gibbs sampling), where the index refers to the Markov step.  If
$\chi^2_{n} > \chi^2_{n-1}$, the algorithm accepts the new parameter
set with probability $e^{-(\chi^2_{n} - \chi^2_{n-1})/2}$.  We 
adopt the medians of the accepted steps as the best-fit set, and we
calculate both upper and lower standard errors using the
16th and 84th percentile values. The error of the $2.8<R_p<4$ bin 
is calculated from the standard deviation of $1/\kappa_3$, i.e. 
the relative occurrence of the 2.8-4\rearth{} bin with respect to 
the 2-2.8\rearth{} bin.

\subsection{Iterative Simulation (IS)}
\label{sec:IS}  
\begin{figure}[h]
\centerline{\includegraphics[scale=0.55]{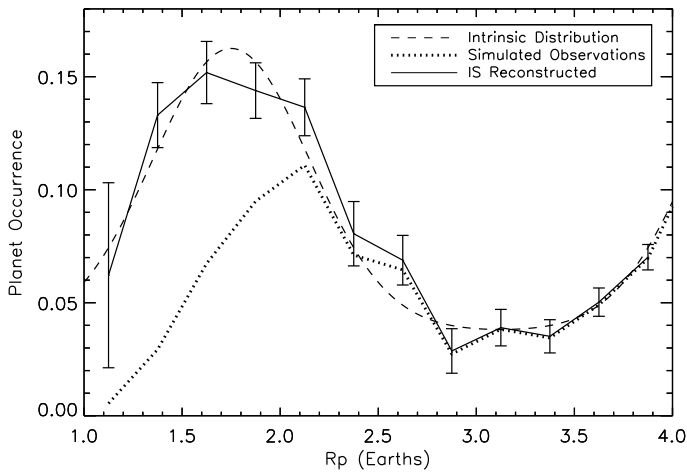}}
\caption{A test case demonstating the recovery of a known, artificial
  radius distribution (dashed line) using the method of iterative
  simulation.  The dotted line represents the simulated observations,
  which consist of 364 detections, and the solid line is the
  reconstructed distribution (here binned for display).  Error
  bars are determined from 25 bootstrapped simulations.}
\label{fig:IS}
\end{figure}

We use the method of sequential Monte Carlo with bootstrap filter described
in \citet{Olivier2007}, which we hereafter refer to as Iterative Simulation (IS),
to infer the intrinsic planet radius distribution without resorting to binning.  
In this technique we first generate a trial population of planets by 
simulating detections (\S \ref{sec:geneq}).  The radii of
these simulated detections are then replaced by actual KOIs, and the process
repeats until the simulated detections converge on the observations. 
Convergence is established when the radius distribution of 
our simulated detected planets (see dotted line in Figure~\ref{fig:IS}) matches the
observed 430KOI distribution. The radius distribution of the trial population then reflects that of
the intrinsic population of planets.  With a sufficiently large trial
population, the resolution of the radius distribution is limited only
by the amount of information in the observations (i.e. KOIs). 
In principle the periods can also be replaced by actual KOI values, but 
we instead choose to fix the distribution of period values 
using Eq.~\ref{eq:P} and $\alpha$, determined from our MCMC analysis. Since 
we find an excellent $\alpha$ fit to the period distribution (\S \ref{sec:intperiod}) over 
$20<P<200$, applying this distribution instead reduces additional noise in our
measurement.

Our IS simulates $10^6$ transiting planet-star pairs, drawing stars
from the selected catalog with replacement (\S\ref{sec:geneq}).
Eccentricities and arguments of periastron of trial planets are drawn
from Rayleigh and uniform distributions, respectively, and periods are drawn from a
power-law with the index of the best-fit MCMC model (\S
\ref{sec:mcmc}), $\alpha=-0.04$.  Planet radii are initially drawn
from a uniform distirbution over 0.5--6\rearth{}.  Detections are
simulated as described in \S \ref{sec:geneq}.  We randomly
replace the radii of all simulated detected planets with values drawn
from the ``master'' radius distribution (\S\ref{sec:plerr}), 
after correcting for false positives using
the rates in Table 1 of \cite{Fressin2013}.  
For a discussion of our false positive treatment, see
\S\ref{sec:sensitivity}. We then redraw new values
for all the other planet parameters besides radius and period, and
reshuffle the planets among the stars.  We repeat this process until
acceptable convergence is achieved, usually within 100 iterations.  At
this point, the trial population is used to calculate the intrinsic
radius distribution. 

Errors are calculated by constructing 50 bootstrapped samples of the
detected planet catalog.  The size of each sample is a random Poisson
deviate with expectation equal to the size of the actual sample.  The
bootstrapped samples are drawn with replacement from the actual
KOI sample.  Planets are randomly removed according to the false
positive probabilities of \citet{Fressin2013} and then intrinsic
radius distributions are calculated as in \S \ref{sec:plerr}.
For our bootstrapped samples we make the false positive correction 
before constructing the much larger intrinsic distribution to 
capture the contribution of false positives to the ``noisiness'' 
of each bootstrapped sample. We analyze each sample using the IS technique 
and compute standard deviations of the ensemble of bootstrapped
planet populations to represent 1$\sigma$ uncertainties.

Figure~\ref{fig:IS} shows the results of an artificial test case of
the reconstruction of a planet radius distribution using the IS technique.  
The intrinsic distribution (dashed line) is the sum of a Gaussian plus a
rising slope.  The dotted line is the distribution of 364 simulated 
observations, which is similar in scale to our
430KOI sample.  We apply the IS technique on these simulated
observations to recover the actual distribution: the result is plotted
as the solid line, with error bars determined from 25 bootstrapped
runs.  The ability of IS to reconstruct an intrinsic distribution is
limited by the information available in any region of a distribution,
i.e. it will fail where the number of planets or rate of detection is
too low.  In Fig. \ref{fig:IS}, errors or large uncertainties appear
in the reconstructed at $R_p \sim 1$\rearth{} where the detection
efficiency is low.  Also, in this simple demonstration we ignore the
effect of planet radius errors.  Adding planet errors tends to broaden
and smooth features.

\subsection{Calculation of Occurrence}
\label{sec:calocc}

After obtaining best-fit distributions from our MCMC and IS
analyses, we calculate the rate of planet occurrence as a function
of $P$ and $R_p$. We generate $10^6$ mock star-planet pairs and the corresponding
simulated detections (\S \ref{sec:geneq}). These planets are binned in
a logarithmic grid of $P$ (index $i$) and $R_p$ (index $j$). The
occurrence $f(i,j)$ in the bin $(i,j)$ is then
\begin{equation}
f(i,j) = \frac{K_{ij}}{N_*}\frac{N_{*,S}}{S},
\label{eq:ISocc}
\end{equation}
where $K_{ij}$ is the false positive-corrected \citep[Table
1,][]{Fressin2013} number of KOIs falling into the bin $(i,j)$,
$N_{*}$ (=76,711) is the total number of \kep{} target stars in
the Solar76k sample, $N_{*,S} (=10^6)$ is the total number of
mock pairs, and $S$ is the number of simulated detections.  
The ratio $S/N_{*,S}$, the fraction of mock pairs that should be detected, 
and is the product of the geometric factor $R_*/a$ as well as the detection 
completeness in bin $(i,j)$, hereafter known as $C(i,j)$ (see \S
\ref{sec:completeness}).  We then sum over all bins to obtain the total 
occurrence, $f$.

To demonstrate the importance of accounting for radius errors 
(\S\ref{sec:plerr}) when calculating occurrence, we also conduct 
a separate analysis which excludes radius errors. Observed planets are 
binned as above and we calculate the occurrence of each bin, $f$ according to:
\begin{equation}
f(P_i,R_i) = \frac{1}{N_*}\sum\limits_{k}^{n_p(i,j)}\left(\frac{a_k}{R_{*,k}}\frac{1}{C(i,j)}\right)
\label{eq:occtw}
\end{equation}
Where $C(i,j)$ is the average completeness of the bin (see \S
\ref{sec:completeness}), $n_p(i,j)$ is the number of planets in bin $(i,j)$, 
$N_*$ (=76,711) is the total number of stars in the sample and $a_k/R_{*,k}$ is 
the geometric correction factor for planet $k$. 

\section{The Primordial Population of \kep{} Planets}

\subsection{Completeness}
\label{sec:completeness}

For a transit survey, completeness is the fraction of transiting
planets of a given $P$ and $R_p$ that are actually
detected, i.e. not including the geometric transit probability. 
Accurately capturing the dependence of completeness
on $R_p$ and $P$ is crucial to a robust determination of planet
occurrence.

\begin{figure}
\centerline{\includegraphics[scale=0.55]{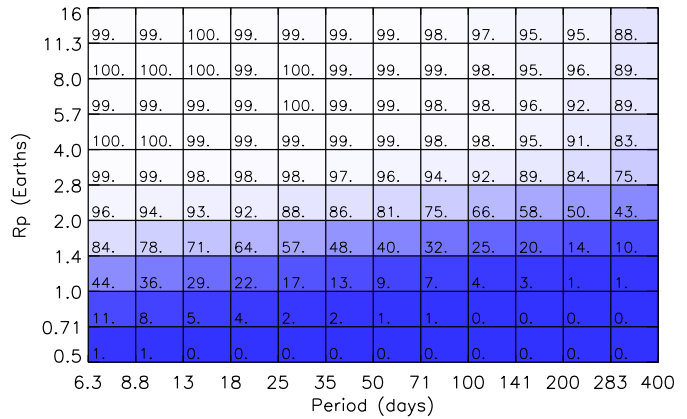}}
\caption{Completeness values for \kep{} planet detection around stars 
  in the Solar76k catalog. The numbers within each grid cell indicate 
  the completeness percentage, and each grid has been colour coded from low
  (blue) to high (white) completeness.}
\label{fig:sil_compl}
\end{figure}

We emphasize that survey completeness depends not only on
the properties of the planet host stars, but also on the stellar
and noise properties of the entire catalog.  We calculate
the completeness $C(i,j)$ of bin $(i,j)$ by inserting $2\times
10^4$ planets randomly distributed around the Solar76k
sample of stars. The fraction of ``detected'' planets, modulo the
transit probability factor, yields the completeness in this bin.
The results are displayed in Figure~\ref{fig:sil_compl}.

Figure~\ref{fig:sil_compl} shows that \kep{} completeness is nearly 100\% 
for planets larger than
Neptune (3.8\rearth{}) and for nearly the full range of periods shown
here. This falls rapidly beyond $P \sim $ 500~d (not shown)
since some systems no longer have the required three
transits during the four-year \kep{} mission. The completeness
drops rapidly with decreasing radius where $R<2$\rearth{}:
Earth-sized planets are readily detected by \kep{} only if they have
orbital periods of a few days, and beyond $P=200$~d, even the
completeness of $2 R_\oplus$ planets falls below 50\%. For these
reasons we restrict our analysis to $P < 200$d.   As true of 
any completeness study, we also note that 
our completeness calculations are based on the SNR criterion as 
stated. For example, if the \kep{} pipeline 
was uniformly missing 25$\%$ of all transiting planets (regardless 
of the SNR), our completeness calculations would not account for these
missing planets.


\subsection{Period Distribution}
\label{sec:intperiod}

Our MCMC study yields a best-fit distribution for the 430KOI sample of
$\alpha=-0.04 \pm 0.09$ (see Eq.~\ref{eq:P}), with a reduced 
chi-squared $\chi^2_{\nu}$ of
1.07. Our value of $\alpha$ is consistent with zero (a flat
logarithmic distribution) within errors, confirming previous
determinations \citep[e.g.,][]{Youdin,Howard2012,Petigura2013,Fressin2013}.

\begin{figure}[h]
\centerline{\includegraphics[scale=0.55]{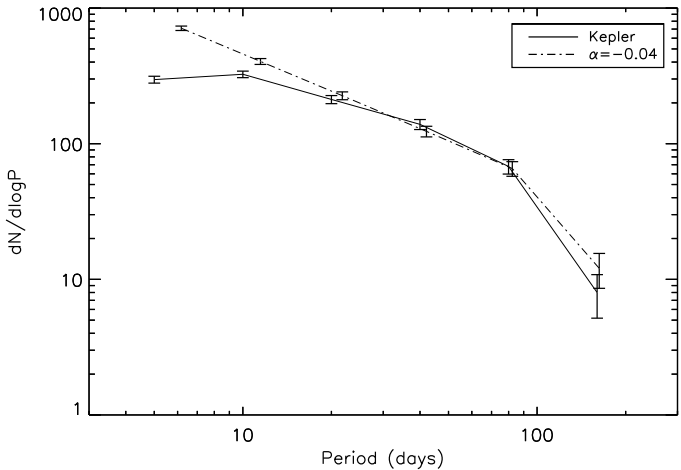}}
\caption{Period distribution of \kep{} small planets. The observed
  distribution (solid line) includes planets inward of $20$~d (i.e. the
  1052KOI sample), while the simulated distribution from the MCMC best
  fit ($\alpha = -0.04$, see Eq.~\ref{eq:P}) is plotted as a
  dashed-dotted curve. This best fit is obtained for planets in the $20 <
  P < 200$~d range, but is extended here to shorter periods to
  demonstrate that the observed population deviates significantly from
  a single power-law shortward of $20$~d. The location of each data point
  corresponds to the lowest period value in the bin, e.g. the first data point is 
  $5<P<10~d$. Slight horizontal offsets have been applied to each curve for clarity.}
\label{fig:period}
\end{figure}

Figure~\ref{fig:period} compares this best-fit period distribution
with the observed sample, using the 4 period bins from Section~\ref{sec:mcmc}
as well as an additional 2 bins to include to include planets inward of
$20$~d, i.e., the 1052KOI sample (\S\ref{sec:catalogs}).
Inside of $P = 20$~d, \kep{}
planets deviate from a simple power-law distribution \citep[also
see][]{Youdin,Howard2012}.

\subsection{Radius Distribution}
\label{sec:radiusdist}

\begin{figure}
\centerline{\includegraphics[scale=0.55]{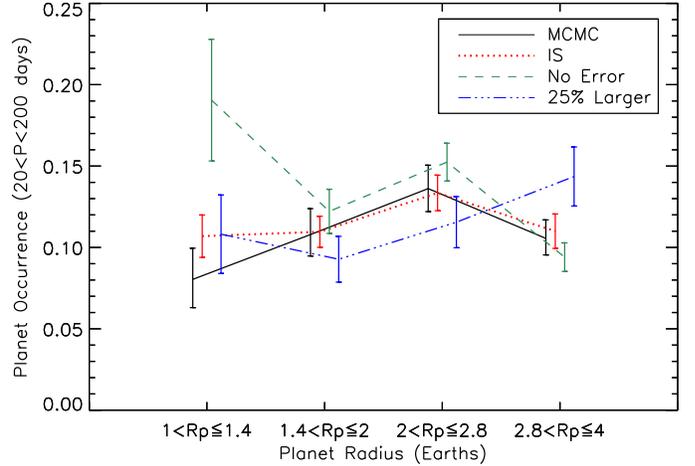}}
\caption{Size distribution of planets between 1 and 4\rearth{},
  obtained using the MCMC (solid black line) and the IS (dotted red
  line) techniques. Both show that planet occurrence peaks in the bin
  2--2.8\rearth{}. Earth-sized planets are less common, 
  though the statistical significance of this result is
  still low. If we assume that the currently determined planet radii
  carry no uncertainty, or that all stars (and hence planets) have
  $25\%$ larger radii than their cataloged values,
  we obtain rather different radius distributions. The error bars for
  the ``No Error'' case account for poisson error only, while for the
  IS and 25$\%$ Larger cases, error bars are calculated from 50 bootstrapped
  simulations of the data (see \S \ref{sec:plerr}). The MCMC error bars are 
  calculated in the standard manner. Planet occurrence at
  each logarithmic radius bin is obtained by summing over all period
  bins.  Slight horizontal offsets have been applied to each curve for clarity.}
\label{fig:raddist}
\end{figure}

Figure~\ref{fig:raddist} displays our best-fit radius distributions
for both the MCMC (solid, black) and IS (dotted, red) techniques,
where we have binned the IS result for ease of comparison.  Both the IS and MCMC
distributions peak at 2--2.8\rearth{} and decrease towards smaller
radii.  We have also plotted two additional distributions in
Figure~\ref{fig:raddist}, a ``No Error'' case (dashed, green)
constructed from Eq.\refnew{eq:occtw} and a ``25$\%$ Larger'' IS case
(dashed-dotted, blue) where it is assumed that both planet and stellar
radii are 25$\%$ larger than their catalog values.


%
 
As we discuss in \S \ref{sec:plerr}, errors in the radius of candidate
\kep{} planets, primarily due to uncertainties in stellar radius, are
large and detection bias against small planets means that a planet's
cataloged radius is likely an underestimate. Comparing our IS and MCMC 
results (which include radius errors) with our ``No Error'' case (which 
doesn't include radius errors) we see that the latter exhibits a significant excess of
1--1.4\rearth{} planets. This is as expected. Correcting for radius error in a
Bayesian way (\S \ref{sec:plerr}) tends to promote small planets to
larger size bins, and de-populates the smallest radius bin. However, planet occurrence of
the two largest bins does not increase significantly since the survey 
completeness is substantially higher in these bins compared to the 
1--1.4\rearth{} bin. We conclude that not accounting for this detection bias 
on radius leads to an erroneously high (by a factor of $\sim 2$)
value of occurrence for the 1--1.4\rearth{} bin.

If we assume the extreme scenario that the true radii of all
stars (and therefore their planets) are 25\% larger than the KOI
values (``25\% Larger'' case in Fig. \ref{fig:raddist}), as is shown
to be the case for at least a subset of the \kep{} stars (\S
\ref{sec:sensitivity}), we observe that the 2--2.8\rearth{} peak is now shifted to
2.8--4\rearth{}. However, we do not see a significant change in
the bin 1--1.4\rearth{} because the depopulation of this region 
(due to increased planet radii) is roughly balanced by a decrease
in completeness as the stars have also become larger. See \S
\ref{sec:sensitivity} for a more detailed discussion.

Our treatment for the radius error is far from
perfect. Some genuinely small planets may, under our procedure, be
wrongly inferred to have larger radii. A better treatment will require
improved errors in stellar/planet radius (see, e.g. \S \ref{sec:improvements}).

There is a small and statistically insignificant discrepancy
between the MCMC and IS results at the smallest bin ($1<R_p<1.4$\rearth{}).  
Since the two methods use
identical input catalogs (\S \ref{sec:catalogs}) and detection
algorithms (\S \ref{sec:geneq}), the difference could be due to the
intrinsic binning in the MCMC method. 

\begin{figure}
\centerline{\includegraphics[scale=0.55]{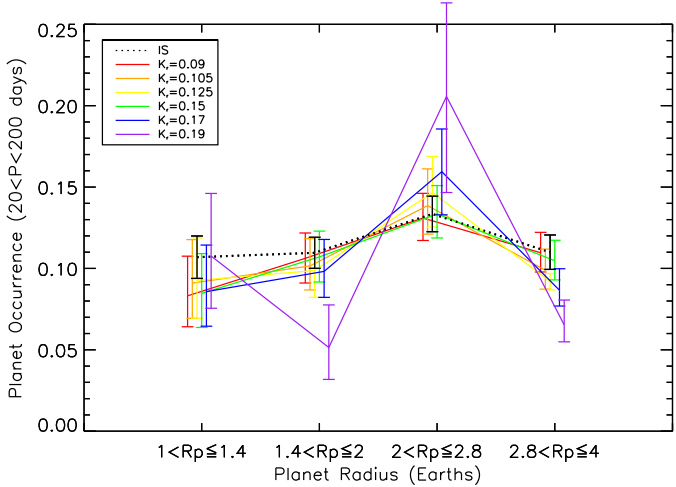}}
\caption{Investigating the effect of binning on the MCMC results. The IS result from Figure~\ref{fig:raddist} is shown as a dotted, black line, 
while our MCMC results using various bin sizes are the coloured curves. Using
Equation~\ref{eq:K}, we vary $K_r$ from 0.09 to 0.19, keeping $K_p$ fixed. 
As can be seen, the results can change noticeably depending on
the binning choice. This illustrates the usefulness of the IS method, which does not 
require any binning. To clarify, we do not change the number of free parameters in our
MCMC analysis (i.e. $\kappa_1, \kappa_2, \kappa_3, \alpha$), merely the size of the
bins used to calculate $\chi^2$ values. Slight horizontal offsets have been added for clarity.}
\label{fig:bintest}
\end{figure}

We test the effect of 
different bin sizes on our MCMC results 
by varying the bin sizes in both radius and period space. Using the
logarithmic binning scheme to describe the width, $w$, of each bin:
\begin{equation}
w(n) = 10^{Kn}
\label{eq:K}
\end{equation}
where $n$ is the bin number and $K$ is a constant, we 
investigate the effect of varying $K$ (and thus bin size) on occurrence.  
When varying period bin size, $K_p$, and keeping the radius bin size, $K_r$, 
constant we find no effect on occurrence, even when $K_p$ is varied by a factor 
of 2. However, when $K_p$ is kept constant and $K_r$ is varied, we do find an affect on  
occurrence rates shown in Figure~\ref{fig:bintest} (where our IS result from 
Figure~\ref{fig:raddist} is plotted as a dotted black line). 
As $K$ is increased (larger bins), we see that the occurrence of $2<R_p<2.8$\rearth{} planets 
increases while the occurrence of $2.8<R_p<4$\rearth{} planets decreases, becoming 
significantly different from our IS result for extreme cases. In contrast, there is no statistically 
significant difference in the smallest bin. We find that 
the error bars increase with more extreme bin sizes, indicating that the MCMC algorithm
has a harder time converging. The default bin sizes for the MCMC result in
Fig.~\ref{fig:raddist} are $K_p$=0.301 and $K_r$=0.150515.

\begin{figure}
\centerline{\includegraphics[scale=0.55]{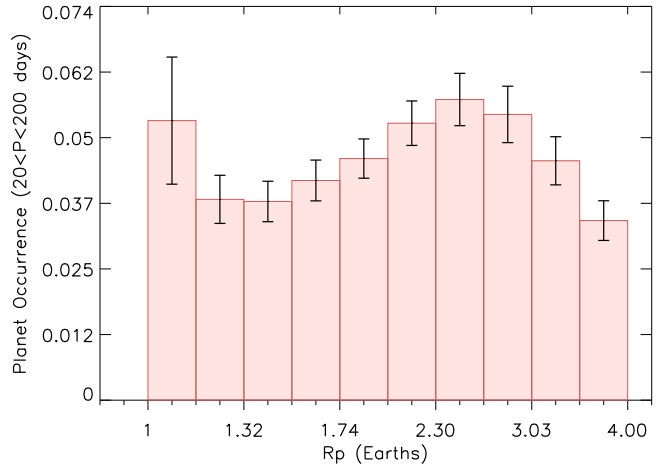}}
\caption{Our IS distribution displayed for a smaller logarithmic 
 bin size. This finer resolution reveals more information about
 the intrinsic distribution, and specifically we see a potential
 rise in the number of 1--1.15\rearth{} planets. This bin has large 
 error however, and thus more statistics are required to confirm this 
 conclusion.}
\label{fig:isnobin}
\end{figure}

Lastly, we display the IS radius distribution for a smaller
logarithmic bin size in Figure~\ref{fig:isnobin}. As explained in 
\S\ref{sec:IS}, since the IS technique requires no binning, the 
resolution of the result is limited only by the data and its errors. 
This improved resolution can reveal finer detail about the intrinsic 
radius distribution. We observe a slight excess of planets in the now 
smallest bin (1--1.15\rearth{}), over that in larger bins, though improved 
statistics are required to confirm this upward turn.  We
expect that, with its independence on binning, the IS technique will
become central to future analysis.

\subsection{Total Occurrence of Small Planets}
\label{sec:totaloccur}
In Table~\ref{tab:occ} we report our estimates for the total planet
occurrence within $20 < P < 200$~d and $1 < R_P < 4$\rearth{}, for
the four curves in Figure~\ref{fig:raddist}. There is excellent
agreement between the MCMC and the IS results. Even cases with
different assumption about the radius error yield statistically
consistent occurrence rates.

The total occurrence rate we calculate here is defined to be the
average number of planets per star
\citep{Youdin,Fressin2013,Petigura2013}. Such a definition ignores
the complication that many of the \kep{} systems are multiple
systems \citep[e.g.][]{Lissauer}. A quantity perhaps more
relevant for studies of planet formation is the occurrence
rate of planetary systems, or the average number of planetary
systems a star has. However, this requires knowledge of the system
architecture, a task not yet attempted.  

Lastly, if we include small planets on orbits interior to $20$~d, the total
occurrence rate is raised to $\sim 66\%$. It will rise by another
$\sim 15\%$ if we include planets larger than $4 R_\oplus$.

\begin{table}[h]
\centering
\caption{Planet Occurrence for $20<P<200$~d, $1<R_p<4$\rearth{}.}
\begin{tabular}{c|c}\hline Technique & Planet Occurrence ($\%$)\\\hline
MCMC                   & $43 \pm 3$ \\
IS                     & $46 \pm 3$ \\
No Error               & $56 \pm 10$ \\
25$\%$ Larger Stars    & $46 \pm 3$ \\
\hline
\end{tabular}
\label{tab:occ}
\end{table}

\subsubsection{Eta-Earth}

We estimate \ee{}, the occurrence of Earth-like planets in the
``habitable zone'' of solar-type stars. By Earth-like, we are
referring to planets between 1--2\rearth{}.  We adopt the inner
and outer boundaries of the habitable zone to be those calculated by
1-D, cloud-free, climate models \citep{Kasting,Kopparapu2013a}.
For a Sun-like star, these boundaries lie at $0.99$ and $1.70$ AU respectively
(or orbital periods of 350 and 810 days); for other stellar spectral
types, the boundaries are as tabulated in \citet{Kopparapu2013a}.

Such a habitable zone, however, lies outside the $200$~d limit of
our study. At these very long periods, the low detection efficiency of
\kep{} engenders inaccuracies in estimating \ee{}. So instead, we have
opted to calculate \ee{} by extrapolation, according to:
\begin{equation}
\eta_{\oplus} = \frac{f}{N_*}\sum\limits_{i=1}^{N_*} h_i
\end{equation}
where $N_*$ (=76,711) is the number of stars in the Solar76k
sample and $h_i$ is the relative occurrence of planets (per star)
within the habitable zone to some reference zone having absolute
occurrence $f$.  We choose this reference zone to be our standard $20 < P <
200$d, $1 < R_p < 4 R_\oplus$ bound, with $f=0.46 \pm 0.03$ (IS
value). We calculate $h_i$ for each star by adopting the IS
radius distribution (Figure~\ref{fig:raddist}) and integrating
the MCMC best-fit power-law (Eq. \ref{eq:P} with $\alpha=-0.04$) over
each star's habitable limits \citep{Kopparapu2013a}.  Finally, we obtain
\begin{equation}
\eta_{\oplus} = 6.4^{+3.4}_{-1.1} \%
\end{equation}
The error is calculated from error propagation of the IS radius
distribution, occurrence of our reference zone, habitable zone limits 
(based on $R_*$, $M_*$, and $T_*$) and $\alpha$.  
This value is consistent within errors with
the analysis done by PHM13 for the same \citet{Kopparapu2013a} limits,
$8.6$\%. The reader is reminded that our calculation of \ee{} is an
extrapolation, and depends crucially on the assumptions made.

\section{Comparison with PHM13}
\label{sec:petcomp}

We now compare our work to PHM13 -- an analysis which
is similar to ours in terms of scope but which obtains their results
of the \kep{} data using the TERRA pipeline \citep{Petigura2013}. The TERRA 
pipeline is an analysis tool independent of the \kep{} project pipeline and its
products on which our work relies.  

\begin{figure}[h]
\centerline{\includegraphics[scale=0.55]{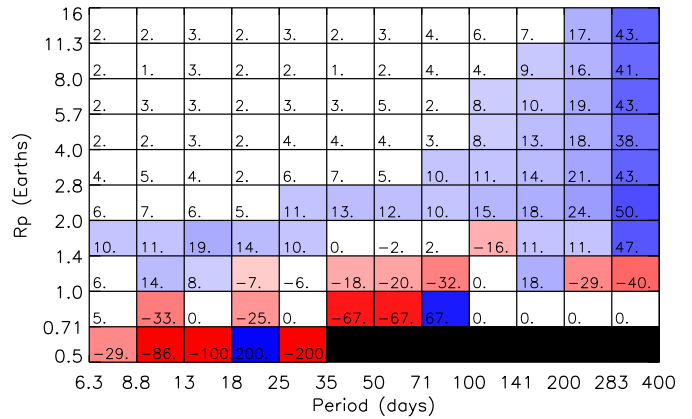}}
\caption{Comparison of completeness values computed with our methods
  to those reported by PHM13 in Figure S11, expressed as a fractional
  difference (i.e. in percentages).  Our values are generated using the ``Best42k'' catalog
  from PHM13 and our detection criteria.  Bins where both our
  completeness and those of PHM13 are zero have been blacked out.}
\label{fig:comp_comp}
\end{figure}

We first compare our estimates of detection completeness $C$ with that
of PHM13.  For this completeness comparison, we re-compute $C$ using the Best42k
stars from PHM13 and compare these results to the values in Figure S11
from PHM13.  We calculate the fractional difference ($2(T-P)/(T+P)$,
where $T$=This Work and $P$=PHM13) and display as percentages in
Figure~\ref{fig:comp_comp}.  
With the exception of a single cell all
values of $C$ in the range $P$ = 20--200~d, and $R_p$= 1--4\rearth{}
are within 20\% of PHM13 values.  This shows that even a comparatively
simple description of \kep{} planet detection can account for most of
the statistics.  The single exception is for the 1--1.4\rearth{} and
71--100~d bin where our estimate of $C$ is 32\% lower than that of
PHM13.  This bin includes the 90~d roll-period of \kep{}.  Large
systematics appear in raw \kep{} lightcurves at this period because
the stars change positions on the detector array.  It might be
expected that the planets with $P$ near 90~d would be more
difficult to detect than our naive criteria and that actual
completeness would be lower.  If the PHM13 values are more realistic,
then the opposite appears to be the case.  Elsewhere in $P$-$R_p$
space our completeness values are slightly and systematically higher
than those of PHM13, and the discrepancy increases with increasing $P$
and decreasing $R_p$.  This is to be expected because PHM13 determine
detection efficiency using actual lightcurves rather than
representations of noise a la CDPP values.  At $P>200$~d our values of
$C$ become significantly higher than PHM13 for nearly all values of
$R_p$.  This discrepancy motivates our restriction to $P < 200$~d.
One possible explanation for this difference is that detection of
signals by phase-folding in the \kep{} detection pipeline becomes
inefficient at long periods.

We note that PHM13 calculates completeness 
based on a finite number ($4 \times 10^4$) of systems of which very few detections 
are in the Earth-sized bins, leading to large counting (Poisson) error in completeness 
values.  Since we use a different approach, simulating large numbers ($2 \times 10^6$ total, 
$2 \times 10^4$ per bin) of test planets and giving high drawing probability 
to Earth-sized planets, we are able to simulate a much larger number of detections 
for each bin and thus have a more precise (although not necessarily more accurate) 
value of completeness.

\begin{figure}
\centerline{\includegraphics[scale=0.55]{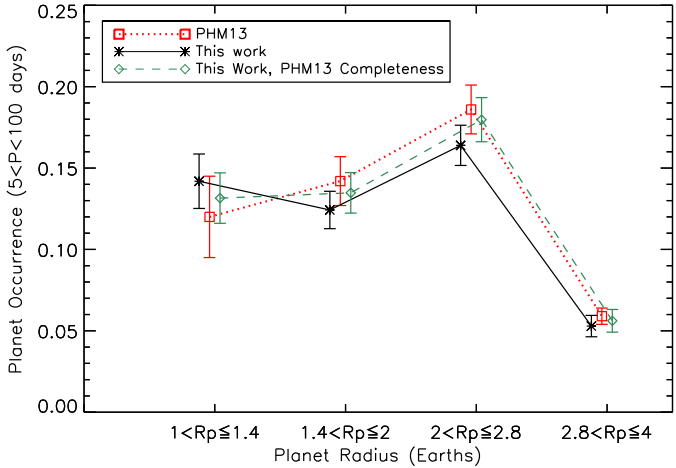}}
\caption{A summary of our comparison to PHM13, plotted in percentages 
  along the y-axis. All curves are constructed using the ``603PHM'' 
  dataset and ``Best42k'' sample. The results
  of PHM13 are plotted as the dotted curve in red with
  squares, ``This work'' uses our own completeness values, and is the black
  solid curve in black with stars, ``This work, PHM13 Completeness'' uses PHM13's Figure
  S11 completeness values and is the green
  dashed curve with diamonds. Slight horizontal offsets have
  been applied to each curve for clarity.}
\label{fig:mastercomp} 
\end{figure}

We next compute the impact of these differences in completeness on
occurrence over $P$=5--100~d, shown in Figure~\ref{fig:mastercomp} (note that the
radius errors of \S~\ref{sec:plerr} have been omitted here). We first 
calculated occurrence using the ``603PHM'' dataset (\S\ref{sec:catalogs}),
Eqn.~\ref{eq:occtw} along with our own detection criteria (\S \ref{sec:geneq}) and completeness
values (calculated from the Best42k sample), shown as the solid black curve in
Figure~\ref{fig:mastercomp}. We then re-calculated planet occurrence
using the 603PHM dataset, Eqn.~\ref{eq:occtw} along with our own detection criteria
but substituting in the completeness values from Figure S11 of PHM13 for
ours, shown as the dashed green line in Fig.~\ref{fig:mastercomp}. Lastly,
the direct results of PHM13 are shown as a dotted, red line. 
The differences between all curves in
Figure~\ref{fig:mastercomp} are small and, within errors, agree with
each other. The residual differences in completeness
seen in Fig.~\ref{fig:comp_comp} do not appear to play a significant role in the
comparative occurrence of our work and that of PHM13 but could be
responsible for some of the minor (and statistically insignificant)
differences that we find. {\it We conclude that simple detection
criteria and noise model can be used in planet occurrence studies to achieve
accurate and precise results}.

Comparing PHM13's intrinsic radius distribution 
(red curve in Figure~\ref{fig:mastercomp}) to our own (black curve in Figure~\ref{fig:raddist}), 
we notice a difference in the occurrence of large ($2<R_p<4$) planets.
Specifically, PHM13 calculated an occurrence of $18.5 \pm 1.5\%$ and $6 \pm 0.5\%$ 
for $2<R_p<2.8$ and $2.8<R_p<4$ planets, respectively, while we find
an occurrence of $13.0 \pm 1.1\%$ and $10.5 \pm 1.0\%$ for the same bins. This results in a statistical 
difference of about $2\%$ and $3\%$ between works for the $2<R_p<2.8$ 
and $2.8<R_p<4$ bins, respectively. 
We see a few possible reasons for this. Table~\ref{tab:comp} displays the major differences 
in raw samples between our work and PHM13, displaying the raw occurrence of large 
($2<R_p<4$\rearth{})
planets for their nominal period ranges, $P>20$~d and $P<20$~d. The bottom two rows of
``This Work'' are empty in Table~\ref{tab:comp} since all of the planets in our sample are $P>20$, 
and thus all the relevant information is present in the first row.  
\begin{table}[h]
\centering
\caption{Differences in raw counts between our work and PHM13}
\begin{tabular}{|c|cc|}\hline Range & This Work & PHM13 \\\hline 
Nominal $P$ range: & & \\
$1<R_p<4$\rearth{} (Total) & 394 & 495 \\
$2<R_p<2.8$\rearth{} & 166 ($42 \%$) & 191 ($39 \%$) \\
$2.8<R_p<4$\rearth{} & 120 ($30 \%$) & 87 ($18 \%$) \\\hline
$P>20$~d: & & \\
$1<R_p<4$\rearth{} (Total) & -- & 201 \\
$2<R_p<2.8$\rearth{} & -- & 89 ($44 \%$) \\
$2.8<R_p<4$\rearth{} & -- & 45 ($22 \%$) \\\hline
$P<20$~d: & & \\
$1<R_p<4$\rearth{} (Total) & -- & 294 \\
$2<R_p<2.8$\rearth{} & -- & 102 (35$\%$)\\
$2.8<R_p<4$\rearth{} & -- & 42 (14$\%$) \\\hline
\end{tabular}
\label{tab:comp}
\end{table}

To summarize Table~\ref{tab:comp}, of the 394
KOIs in our raw planet sample between $1<R_p<4$, 166 ($42 \%$) and 120 ($30 \%$) 
are between $2<R_p<2.8$\rearth{} and $2.8<R_p<4$\rearth{}, respectively, 
while for PHM13, of the 495 KOIs between $1<R_p<4$\rearth{}
only 191 ($39 \%$) only 87 ($18 \%$) fall into the same limits, respectively. 
This is a significantly lower fraction, and it appears that our sample and PHM13's sample 
are markedly different. There are only a couple options available to explain this difference -- 
either a disproportionate number of large 
($2<R_p<4$\rearth{}) planets in our sample are false positives, 
or the two datasets are not subsamples from the overall same population. 

If the two samples are from different populations, this difference we claim is due to the
photoevaporation \citep{Owen2013} of PHM13's close-in planets, 
which acts to convert large ($2<R_p<4$\rearth{})
planets into smaller ones. Splitting the PHM13 dataset into $P>20$~d and $P<20$~d subsets
adds support to this theory, since as we move from small to large periods 
raw occurrence drops proportionally by 
$(44 \% - 35 \%)/ 44 \% \sim 20 \%$ and $(22 \% - 14 \%)/ 22 \% \sim 40 \%$ for 
$2<R_p<2.8$\rearth{} and $2.8<R_p<4$\rearth{}  
planets, respectively.



In explaining the difference between the occurrence of large planets between this work and PHM13, 
one must also consider the improved treatment of
radius errors and updated \citet{Huber2014} stellar parameters used in this work. 
Both tend to increase planet size, pushing smaller planets into larger bins.
Comparing the ``No error'' case to the ``IS'' case in Fig.~\ref{fig:raddist}, we can see that
the former effect increases the occurrence of $2<R_p<2.8$\rearth{} and $2.8<R_p<4$\rearth{} 
planets by about 2$\%$ and 1.5$\%$, respectively. Thus, it appears that 
the discrepancy of our results with the PHM13 in the $2<R_p<2.8$\rearth{} bin can
be explained by our incorporation of radius errors. However this does not 
account for the total discrepancy in the $2.8<R_p<4$\rearth{} bin, leaving about $1.5\%$ of
discrepancy between our work and PHM13. 

A final source of discrepancy in the $2.8<R_p<4$\rearth{} bin between works 
may come from the different stellar
populations used between our work and PHM13. \citet{Mulders2014} found that the occurrence 
of planets is correlated with stellar type, and a quick analysis of the Best42k sample shows that 
$\sim 30 \%$ of the stars used in the PHM13 analysis are outside the $0.8<R_*/R_{\oplus}<1.2$
range. However, it should be noted that \citet{Mulders2014} does not predict
a significant change in occurrence across stellar types for $2.8<R_p<4$\rearth{} planets.

To conclude this comparison, PHM13 reported an occurrence of $37 \pm 3.4\%$ for planets with 
$25<P<200$~d and $1<R_p<4$\rearth{}. Including planets from 
$20 < P < 25$~d raises this value to $42 \pm 3.6\%$.  
So the overall occurrence rates are consistent among
studies that are based on different detection criteria and different
model assumptions. 


\section{Discussion}
\label{sec:discussion}

\subsection{Sensitivities and Systematics}
\label{sec:sensitivity}

To investigate sensitivities to some of the assumptions and parameters
in our analysis, we repeat our IS simulations varying $\sigma_e$ and $\alpha$. First,
we varied the Rayleigh distribution of orbital eccentricities between 0.1
and 0.3, and second varied the value of $\alpha$ between -0.15 and 0.15.  
In all cases we found no significant difference in the radius distributions. We also
investigate our separability assumption (Eqn.~\ref{eq:assumption}) by
splitting our 430KOI dataset into two equal sized subsets corresponding to $20<P<50$ and
$50<P<200$~d planets, and performing an MCMC analysis on each. 
The resulting distributions are consistent with each other as well as with our
main IS and MCMC results, indicating that there is no significant
correlation between planet radius and period in our sample, and
that Eqn.~\ref{eq:assumption} is a reasonable assumption.

In this analysis, we address the fact that many \kep{} planets
have large errors in radius, driven primarily
by uncertainties in the radii of their host stars.
But we have not addressed the issue of systematic errors
in stellar radii.  For example,
stellar effective temperatures based on photometry from KIC
\citep{Brown2011} are systematically $\sim$200~K hotter than more
reliable estimates based on the infrared flux method
\citep{Pinsonneault2012} or spectroscopy \citep{Gaidos2013}.  The
combination of uncertainties in stellar parameters and Malmquist bias
in the magnitude-limited \kep{} target catalog means that the
sample is biased towards the most luminous, hottest, and largest stars
\citep{GaidosMann2013}.  There is increasing evidence that many \kep{}
target stars, including planet hosts, are subgiants \citep[e.g.,
][]{Verner2011,Everett2013,Bastien2014}.  For fixed values of
$R_p/R_*$, systematically larger stellar radii means the planets
are also systematically larger and that the geometric transit
probability is higher than presumed (transit probability depends
inversely on stellar density and hotter, more evolved stars are less
dense).  The detection completeness of small planets is also smaller
than presumed and thus, for fixed number of detections, the occurrence
is higher.

We have explored the possible impact of these effects by assuming
that all stellar radii in our Solar76k catalog, as well as all
the planet radii in the corresponding 450KOI catalog, are 25\% larger
than their nominal values (dashed-dotted blue curve, Fig.\ref{fig:raddist}). 
The distribution differs markedly from our IS and MCMC distributions. 
The peak in the distribution at 2--2.8\rearth{} has shifted towards
larger radii.  Surprisingly, the occurrence of planets with $R_p$ =
1--1.4\rearth{} has not changed.  This is understood. As stellar
radii increase, completeness of small planets decrease leading to an
increase in planet occurrence. However, the number of 1--1.4\rearth{}
planets in our sample also decreases (from 25 to 8 after a 25$\%$
radius increase), reducing the raw planet
occurrence in this radius bin. It appears that these two competing
effects roughly cancel, resulting in no significant change in planet
occurrence of the smallest radius bin. 

We now comment on our treatment of false positives in this analysis. 
Our work makes use of the \citet{Fressin2013} false positive rates based 
on the Q1--Q6 \kep{} data, while other works (e.g. PHM13) use custom 
methods to detect false positives. 
However since we use the latest disposition of the \kep{} catalog in this analysis,
the occurrence of false positives in our Solar76k sample could be significantly
different. We calculate the ratio of false positives (``FP'') vetted by the
\kep{} science team to planet candidates (``CAN'') for the Q1--12, Q1--16 and
cumulative \kep{} catalogs according 
to FP/(FP+CAN). In addition, we organize these false positive ratios by radius, 
using the same radius bins as our analysis. These false positive ratios are shown 
in Table~\ref{tab:fp}, as well as the \citet{Fressin2013} values for reference. 
It should be noted that the fraction of planets yet to be 
dispositioned (calculated according to DISP/(DISP+CAND+FP), where DISP is the 
number of planets yet to be dispositioned) in the Q1--12 and Q1--16 datasets are quite high 
(over 50 $\%$), and may reflect the significant difference between their false positive rates 
and the cumulative KOI dataset rates. 

\begin{table}[h]
\centering
\caption{False positive ratios, calculated according to FP/(FP+CAND)}
\begin{tabular}{c|c}\hline Dataset  & FP$\rightarrow$[1-1.4, 1.4-2, 2-2.8, 2.8-4]\rearth{} \\\hline
\citet{Fressin2013} & 0.088, 0.088, 0.067, 0.067 \\
Q1--12             & 0.37, 0.39, 0.27, 0.24 \\
Q1--16             & 0.28, 0.31, 0.36, 0.53  \\
Cumulative      & 0.22, 0.24, 0.16, 0.19 \\
\hline
\end{tabular}
\label{tab:fp}
\end{table}


We use the false positive fractions in table~\ref{tab:fp} to estimate the uncertainty
of using \citet{Fressin2013} false positive rates in our analysis. 
For Q1--12, Q1--16 and the cumulative list, 
we carry out separate MCMC analyses substituting in the false positive rates from 
Table~\ref{tab:fp}.  Only the cumulative list remains consistent with our main IS and MCMC
results, with overall occurrence rates of the Q1--12, Q1--16 and cumulative datasets
being $0.30 \pm 3\%$, $0.32 \pm 3\%$ and $0.38 \pm 3\%$, respectively, i.e. much lower
occurrence values. 
Taking the standard deviation of each radius bin for our main IS result (Fig.~\ref{fig:raddist})
plus these three new analyses (i.e. using Q1--12, Q1--16 and cumulative false positive rates) 
we estimate the uncertainty in our false 
positive rates on the occurrence of planets for $1<R_p<1.4$, $1.4<R_p<2$, $2<R_p<2.8$, and
$2.8<R_p<4$\rearth{} to be 2.1$\%$, 1.6$\%$, 1.8$\%$ and 2.4$\%$, respectively.


\subsection{Astrophysical and Astrobiological Implications}
The period distribution of \kep{} small planets contains two distinct
parts. The first is a rise from  $\sim$ 5--10~d 
\citep[e.g.][]{Youdin,Howard2012}, the second is a
logarithmically flat distribution extending from $\sim 10$~d out to
at least 200~d \citep[Fig. \ref{fig:period}
here,][]{Petigura2013,Fressin2013}. The origin of both features are
unclear. But we speculate on one origin for the logarithmically flat feature.
Imagine a set of planetary systems comprised of
closely-packed, equal-mass planets. Dynamical stability requires that
neighbouring planets be spaced apart by more than a few Hill radii
\citep{Chambers,SmithLissauer}. Since the Hill radius scales linearly
with orbital semi-major axis, this means the separation between
neighbouring planets grows linearly with their orbital
span. This would then translate into a period
distribution that is flat in logarithmic period. In other words, it is
possible that most or all of our planets are actually in multiple
systems, and that the flat feature is a result of the stability
requirement. Although \citet{Fang2012} have quoted that $75 - 80 \%$
of planetary systems have one or two planets with orbital periods less than 
200~d (suggesting that most systems are not close to the Hill stability limit), this
result is based on the occurrence of {\it observed} systems. Most Earth-sized
planets (or smaller) around Sun-like stars are undetectable by \kep{}, and
it is possible that the multiplicity of such systems are much higher than
we currently believe due to these unseen planets.

Alternatively, the flat feature can arise from the primordial mass
distribution in the disk. Assuming that all planets have comparable
masses, are in multiple systems, and are
formed where they are found today, a logarithmically flat
spacing would suggest that the disk surface density $\Sigma$ scales
with the orbital separation $a$ as,
\begin{equation}
\Sigma \propto a^{-2}\, .
\end{equation}
This is not vastly different from the theoretical MMSN profile:
$\Sigma \propto a^{-3/2}$ \citep{Hayashi,Weidenschilling}, a
useful benchmark to study proto-planetary disks.

The radius distribution is equally intriguing. The radius of a Earth- to Neptune-sized
planet mostly reflects the expanse of its hydrogen envelope \citep{Wolfgang2014}.  By
focusing on planets outward of $20$~d, we discard candidates that
may have had their atmospheres eroded by stellar irradiation \citet{Owen2013}. The
distribution shown in Fig.\ref{fig:raddist} is therefore likely
``primordial''. Compared to planets inward of $10$~d that have radii
$\leq 1.5 R_\oplus$, this ``primordial'' population appears to prefer a 
size of $\sim 2.5R_\oplus$. Such a size corresponds to a fractional mass in the
hydrogen envelope of $\sim 1\%$ \citep[assuming a rocky core roughly 
in the $10M_{\oplus}$ range, see, e.g.][]{WuLithwick}. What
is the reason behind this preferrence for $1\%$? A planet embedded in
a proto-planetary disk can accrete a hydro-static
atmosphere. \citet{Rafikov} calculated that this atmosphere has a mass
of a few $M_\oplus$ for a $10 M_\oplus$ planet at $0.1$ AU in a MMSN
disk. This lies much above the $1\%$ value but it depends on disk
parameters and its evolution history.  In future works, the observed
radius distribution should be used to decipher formation history.

Moreover, the gradual decline toward smaller
sizes in logarithmic space has implication for the formation of
bare-core planets, the norm in the inner Solar
system. Our terrestrial planets are thought to have formed in a
gas-free environment by conglomeration of solid materials. The
relative shortage of bare-core planets may suggest that the 
observed \kep{} planets may have followed different formation path 
than that of the terrestrial planets.

Lastly, we turn to the issue of \ee{}. We calculate \ee{} more out of
respect for tradition than with any conviction that there is
additional accuracy to be assigned to our calculation.  The limits of
the habitable zone depend on important assumptions regarding the
climate state of Earth-like planets \citep{Kopparapu2013a}, mass
\citep{Kopparapu2014}, and the composition of the atmosphere
\citep{Pierrehumbert2011}.  Nevertheless, the search for life
elsewhere can take heart in the fact that multiple investigations
point to an occurrence of Earth-size planets in habitable zones of
$\mathcal{O}(0.1)$ or more.  Indeed, studies of M dwarfs suggest that
\ee{} $\sim 0.5$ \citep{Bonfils2013,Kopparapu2013b,Gaidos2013}.  M
dwarfs comprise about 70\% of all stars and hence weigh heavily in the
census for Earth-like planets.

\subsection{Improvements in Occurrence will Happen}
\label{sec:improvements}
The errors associated with most \kep{} planets are dominated by
the uncertainty in the parameters of their host stars. Thus, in 
order to improve planet occurrence calculations for the future we 
must first understand \kep{} stars better. 
The {\it Gaia} (Global Astrometric Interferometer for Astrophysics)
mission, launched in December 2013, will measure the parallaxes of 1
billion stars in the local group with accuracies approaching 10
$\mu$as, as well as obtain multi-band photometry measurements
\citep{deBruijne2012}. \citet{Liu2012} estimate that for stars in the
KIC, {\it Gaia} will be able to estimate $T_{\rm eff}$ to 1\%, $\log
g$ to within 0.1-0.2 dex, and [Fe/H] to within 0.1-0.2 dex.  The
combinations of these data should dramatically improve our knowledge
of the properties of \kep{} target stars and hence reconstructions of
the \kep{} planet population.

Other advances include improved maps of interstellar reddening in the
\kep{} field based on the colors of oscillating red giants with
established properties, as well as WISE infrared photometry
\citep{Huber2014}.  The advent of multiplexed, multi-object
spectrographs capable of simultaneously measuring thousands of stars
\citep{Hill2010} should, combined with {\it Gaia} parallaxes, allow
stellar parameter estimation with unprecedented
scale and precision. In addition, measurement of photometric noise due
to stellar granulation (``flicker'') is a promising technique for
estimating the $\log g$ and hence radius of bright \kep{} stars to
within 0.1-0.2 dex \citep{Bastien2014}, although its calibration
and applicability to fainter \kep{} stars -- the majority of targets,
with lower photometric precision -- remains to be seen.

\section{Conclusions}
In this work we have developed a population simulator to
extract the underlying period and radius distributions of Earth- to Neptune-sized
planets detected by \kep{}. We focus on a ``primordial'' population of
planets outside $20$~d to exclude the impact of, e.g. photoevaporation.
We find that the adoption of a simple model of photometric noise and
transit signal detection allow us to accurately estimate the 
survey completeness of \kep{}.  We have accounted
for radius errors in our analysis, and have found that doing so is
important for reconstructing the intrinsic radius disitribution. We
apply the iterative simulation technique to reconstruct the planet distribution
with radius. This does not require binning and allows radius errors to be
readily accounted for. Lastly, we are the first to use
the updated Huber et al. 2014 parameters along with all 16 quarters of
\kep{} data, representing the most up to date analysis. The main
results are as follows: 
\begin{enumerate}

\item The distribution of planets with $20 < P < 200$ days is roughly uniform with
logarithmic period (power-law index $\alpha = -0.04 \pm 0.09$).

\item The (likely primordial) radius distribution for \kep{} planets
  with $20<P<200$d peaks in the radius bin 2--2.8\rearth{}. 
    
\item The overall occurrence of planets within $20<P<200$~d and
  $1<R_p<4$$R_\oplus$ is $46\% \pm 3\%$. This represents the average
  number of planets per solar-type star in the \kep{} field.

\item Extrapolating our radius and period distributions out to the
  habitable
  zone for solar-type stars, we find $\eta_{\oplus}=6.4^{+3.4}_{-1.1}\%$.\\

\item While our results confirm those from earlier studies, 
 there is a discrepancy in the occurrence of planets for $2.8<R_p<4$\rearth{} planets  
  between our work ($10.5 \pm 1.0\%$) and PHM13 ($6.0 \pm 0.5 \%$). Our incorporation of 
  radius errors and updated \citet{Huber2014} stellar parameters account for about half of this 
  discrepancy, while the difference in raw samples account for the remainder. 
  PHM13 includes $P<20$~d planets into their analysis which likely contains 
  photoevaporated planets (see Table~\ref{tab:comp}), decreasing the overall occurrence
  in the $2.8<R_p<4$\rearth{} bin. 
  We claim that the increase in the occurrence of $2.8<R_p<4$\rearth{} planets in our analysis 
  is due to the exclusion of planets altered by proximity to their host stars. 
  
 \item In a detailed comparison with PHM13 we find that using CDPP
 values can effectively reproduce the detection completeness found by 
 the more sophisticated analysis of PHM13.

\item Large radius errors are present in the \kep{} data, and failing
  to account for these properly can lead to a different radius
  distribution. Specifically, this tends to result in a large excess
  of earth-sized planets. Increasing the size of \kep{} stars by $25 \%$
  increases the frequency of large planets while keeping the occurrence
  of small planets roughly constant. Many stellar radii in the \kep{}
  catalog are suspected to be underestimated, and GAIA will improve
  these stellar radius errors and resolve this issue.

\end{enumerate}

\acknowledgments

The authors would like to thank the referee, Erik Petigura for his insightful comments
which helped to improve this manuscript.
This research uses data collected by the \kep{} mission, as well as
the NASA Exoplanet Archive.  This research is supported by an NSERC
CGS M award to AS, an NSERC discovery grant to WYQ. EG acknowledges
support by NASA grants NNX10AQ36G and NNX11AC33G.

\bibliographystyle{apj}
\bibliography{paperref}

\begin{table}
\centering
\caption{Sample List of 430KOIs and parameters}
\begin{tabularx}{440pt}{cccccccccc}\hline\hline KOI & $P$ & $R_p/R_*$ & $R_*$ & $\sigma_{R_*}$ (+) & $\sigma_{R_*}$ (-) & $R_p$ & $\sigma_{R_p}$ (+) & $\sigma_{R_p}$ (-) & SNR \\
 & (d) & & ($R_{\odot}$) & ($R_{\odot}$) & ($R_{\odot}$) & ($R_{\oplus}$) & ($R_{\oplus}$) & ($R_{\oplus}$) & \\\hline

K00435.05    &  62.3026  &  0.0272600  &   0.832156  &   0.349945 &   0.0623970   &   2.47640  &   0.834870   &   1.32175   &   31.7800 \\
K02289.02    &  20.0984  &  0.0117000   &   1.13947   &  0.705986   &  0.133112    &  1.45539   &   1.62602    &  1.85152    &  16.8800 \\
K00880.01    &  26.4429  &  0.0404000   &  0.928228  &   0.382174  &  0.0842040   &   4.09379   &  0.374281   &   1.68616   &   73.3000 \\ 
K00880.02    &  51.5300  &  0.0540400  &   0.928228  &   0.382174  &  0.0842040   &   5.47595   &  0.497559   &   2.25476   &   153.100 \\
K04150.01    &  31.3352  &  0.0152000  &    1.17072   &  0.559907   &  0.145652   &   1.94262    &  3.05132    &  3.18046    &  14.0000 \\\hline
\end{tabularx}
\label{tab:430KOI}
\tablecomments{The radii of these planets have been updated with Huber et al.,
(2014) parameters where applicable.  Table~\ref{tab:430KOI} is published in its entirety in the
electronic edition of the {\it Astrophysical Journal}.  A portion is shown here for
guidance regarding its form and content.}
\end{table}

\end{document}